\documentclass[final,preprintnumbers]{revtex4}

 \usepackage{amsmath} 
\usepackage{amsfonts} 
\usepackage{amssymb}
\usepackage{graphicx} 
\makeatletter
\DeclareRobustCommand\MakeTextUppercase{%
  \@uclcnotmath{\def\i{I}\def\j{J}}{##1##2}\large}
\DeclareRobustCommand\MakeTextLowercase{%
  \@uclcnotmath{}{##2##1}\lowercase}
\protected@edef\MakeTextLowercase#1{\MakeTextLowercase{#1}}
\let\ProvidesPackage\ProvidesPackage@latex
\let\ProcessOptions\ProcessOptions@latex
\let\DeclareOption\DeclareOption@latex
\expandafter
\let\csname MakeUppercase \expandafter\endcsname
    \csname MakeTextUppercase \endcsname
\expandafter
\let\csname MakeLowercase \expandafter\endcsname
    \csname MakeTextLowercase \endcsname
\appdef\class@documenthook{%
 \switch@longtable
}%
\makeatother

\makeatletter
\relax
  \def\ps@headings{%
      \let\@oddfoot\@empty\let\@evenfoot\@empty
      \def\@evenhead{\thepage\hfil\slshape\leftmark}%
      \def\@oddhead{{\slshape\rightmark}\hfil\thepage}%
      \let\@mkboth\markboth
    \def\sectionmark##1{%
      \markboth {%
        \ifnum \c@secnumdepth >\z@
          \thesection\quad
        \fi
        ##1}{}}%
    \def\subsectionmark##1{%
      \markright {%
        \ifnum \c@secnumdepth >\@ne
          \thesubsection\quad
        \fi
        ##1}}}%
\def\ps@myheadings{%
    \let\@oddfoot\@empty\let\@evenfoot\@empty
    \def\@evenhead{\thepage\hfil\slshape\leftmark}%
    \def\@oddhead{{\slshape\rightmark}\hfil\thepage}%
    \let\@mkboth\@gobbletwo
    \let\sectionmark\@gobble
    \let\subsectionmark\@gobble
    }%
\def\ps@article{%
    \@provide\@evenhead{\let\\\heading@cr\thepage\quad\checkindate\hfil{\leftmark}}%
    \@provide\@oddhead{\let\\\heading@cr{\rightmark}\hfil\checkindate\quad\thepage}%
    \@provide\@oddfoot{}%
    \@provide\@evenfoot{}%
    \let\@mkboth\markboth
  \let\sectionmark\@gobble
  \let\subsectionmark\@gobble
}%
\def\ps@article@final{%
    \@provide\@evenhead{\let\\\heading@cr\thepage\quad\checkindate\hfil{\leftmark}}%
    \@provide\@oddhead{\let\\\heading@cr{\rightmark}\hfil\checkindate\quad\thepage}%
    \@provide\@oddfoot{}%
    \@provide\@evenfoot{}%
    \let\@mkboth\markboth
    \def\sectionmark##1{%
      \markboth{%
        \@ifnum{\c@secnumdepth >\z@}{\thesection\hskip 1em\relax}{}%
         ##1%
       }{}%
    }%
    \def\subsectionmark##1{%
      \markright {%
        \@ifnum{\c@secnumdepth >\@ne}{\thesubsection\hskip 1em\relax}{}%
         ##1%
      }%
    }%
}%
\makeatother

\makeatletter
\def\bibliographystyle{\def\@bibstyle}%
\def\bibsection{%
  \let\@hangfroms@section\@hang@froms
  \section*{\large\refname}%
  \@nobreaktrue
}%
\makeatother
\renewcommand\refname{References}
\begin{document}
\preprint{MIT-CTP/4358}
%\title{Directed Zero Point Motion}
\title{A Long View of Particle Physics\footnote{Rapporteur talk at the 25$^{th}$ Solvay Conference on Physics, ``Theory of the Quantum World'', October 2011.  This write-up closely follows the lecture as delivered, with one exception, explained in a note at the end.}}

\author{Frank Wilczek}
\vspace*{.2in}
\affiliation{Center for Theoretical Physics \\
Department of Physics, Massachusetts Institute of Technology\\
Cambridge Massachusetts 02139 USA}
%\preprint{MIT/CTP-3847}
\vspace*{.3in}

\begin{abstract}
2011 marked the hundredth anniversary both of the famous Solvay conferences, and of the Geiger-Marsden experiment that launched the modern understanding of subatomic structure.   I was asked to survey the status and prospects of particle physics for the anniversary Solvay conference, with appropriate perspective.   This is my attempt.   
\end{abstract}

\maketitle

\bigskip

%%%%%%%%%%%%%

\section{Origins: Understanding Matter}

%%%%%%%%%%%%%%%%%%%%%

\bigskip

The intellectual origin of particle physics is quite straightforward: It arose out of the program of understanding the physical world through ``Analysis and Synthesis'' (Newton), or in modern terminology Reductionism.  More specifically, it arose from the program of studying the smallest building blocks of matter and their properties, in the hope that those building blocks would obey simple laws, from which the nature and behavior of larger bodies could be inferred mathematically.   

The success of that program was by no means logically guaranteed.   Indeed, I think it is fair to say that only in the last hundred years has its slow-ripening fruit matured.   But by now the strangeness, beauty, and richness of that fruit has far exceeded any reasonable, or even mystic, expectation.  We out-Pythagoras Pythagoras.   

Perhaps the most crucial development occurred almost precisely 100 years ago.   I refer of course to the rapid sequence of events beginning with the Geiger-Marsden experiment (1911), advancing with Rutherford's interpretation of that result in terms of electromagnetic ``Rutherford scattering'' from heavy atomic nuclei in solar-system-like atoms (1912), and culminating in Bohr's inspired infusion of quantum ideas into subatomic dynamics (1913).   

After the quantitative success of Bohr's model for hydrogen, and its many other semi-quantitative and qualitative successes, no doubt could remain about its central message: Atoms are held together by electromagnetic forces acting between small, massive nuclei and much lighter electrons, subject to rules of quantization.   It was the work of a generation to create a physically satisfactory, mathematically coherent discipline of quantum dynamics.  After more than a decade of indecisive skirmishing, breakthroughs by Heisenberg (1925), Schr\"odinger (1926), and others of their storied contemporaries rapidly established the basic outlines of quantum theory that we still recognize today.   

The atomic nucleus was at first a tiny black box and a source of bizarre surprises, including notably the various emanations of radioactivity and the need for new binding forces.   Chadwick's discovery of the neutron (1932), and the application of quantum principles, supported the rapid construction of a useful phenomenological description of nuclear phenomena, that we still use today.   

With these achievements the ``practical'' goal of reductionism, for ordinary matter under ordinary conditions, was attained.  An adequate foundation for condensed matter physics, materials science, chemistry, and (presumably) biology was in place.  That foundation remains firm.   

Though in some sense the original goal had been achieved, the intellectual situation was far from satisfactory.    Theoretical nuclear physics, in particular, was a semi-empirical enterprise.  It had not engendered governing equations or principles worthy to stand beside general relativity and covariance, or Maxwell's equations and gauge invariance.  As we now know, in many ways the story was just beginning.

\bigskip

%%%%%%%%%%%%%

\section{Phenomena: New Questions and Surprising Answers}

%%%%%%%%%%%%%%%%%%%%%

\bigskip

The discovery of antimatter, and the successful measurement and calculation of effects of virtual particles, vindicated a radically conservative attitude toward the basic principles of quantum field theory.   So did the experimental discovery of Yukawa's $\pi$ mesons, posited to explain nuclear binding forces.   

On the other hand many ``gratuitous'' new particles, starting with muons and $K$ mesons, spawned a complex of flavor problems, that are still very much with us.   And the complexity of nuclear forces, crowned by discovery of hordes of strongly interacting resonances, seemed to challenge the whole framework of quantum field theory.  

{\it Symmetry}, harnessed to radically conservative quantum field theory, proved to be the most reliable and fruitful guide to decoding Nature's principles.  

\bigskip

Universality and the $V-A$ structure of weak interactions led to the $SU(2)\times U(1)$ electroweak theory.  Besides extending the gauge principle, this theory introduced two major dynamical innovations.   One is the fundamental significance of fermion chirality.   That concept erupted from the discovery of parity violation. The other is gauged spontaneously broken symmetry.  That mechanism for generating gauge boson masses was implicit in the London-Landau-Ginzburg treatment of superconductivity, but the emphases of superconductivity theory were quite different.  To recognize the massive photon interpretation of quanta {\it inside\/} the superconductor, and to give that interpretation a firm foundation in relativistic quantum field theory, were major achievements.   

Patterns among resonances and Bjorken scaling led to the $SU(3)$ color theory of the strong interaction, quantum chromodynamics (QCD).  Besides extending the gauge principle, this theory introduced two additional major dynamical innovations.     One is the confinement phenomenon.   It was shocking, in the early days of the quark model, to contemplate the existence of elementary quanta that do not appear in the physical spectrum.  What could such ``existence'', which explicitly rejects physical existence, possibly mean?   Yet we've come not just to live with confinement, but to understand that it is a natural consequence of gauge symmetry. ({\it Deconfinement\/} is the subtler case!)   The other is asymptotic freedom, which both opens ultra-high-energy processes in cosmology and at accelerators to quantitative treatment, and enables a rigorous, completely nonperturbative approach to calculation of the spectrum, by means of a convergent discretization.

Both theories are, to an extraordinary degree, {\it embodiments\/} of ideal mathematical symmetry.   For example, color gluons were introduced specifically to implement gauge covariant parallel transport, and their properties were derived without ambiguity from purely conceptual considerations, prior to any direct experimental evidence for them.

\bigskip

These two theories together constitute the Standard Model.   The Standard Model overcomes the most unsatisfactory feature of the earlier, semi-phenomenological nuclear theory: It {\it is\/} based on equations worthy to stand beside Einstein's general relativity and Maxwell's electrodynamics.   

With the mature result of reductionism before us, we can consider an important philosophical question: {\it Why\/} does reductionism work?  Several specific aspects of physical theory underpin the success of that approach.   Because of symmetry under translations of time and space, we can infer the laws through repeatable experiments, compiling work done over generations all over the world.   Locality of the laws is also crucial, both to make repeatable experiments a practical possibility, and to guarantee that having completed the Analysis, we can perform the Synthesis.   Relativistic quantum field theory explains the existence of many identical particles, as products of a common field, and thus supplies our elementary building-blocks.  More subtle, but also crucial, is what I call {\it quantum censorship}: the feature of quantum physics, that complex systems can look like ideally simple ones, if we probe them only below their energy gap.   Because the success of reductionism in fundamental physics depends on such specific, non-trivial features of physical law, we should not take the utility of that approach in other domains for granted.

%%%%%%%%%%%%%

\section{Questions That the Standard Model Begs}

%%%%%%%%%%%%%%%%%%%%%

\bigskip

\subsection{Questions from the Core}

The core of the Standard Model -- i.e. the part embodying gauge symmetry -- seems near to perfection.  For just that reason, its remaining flaws stand in sharp relief.  Most obviously, it has more moving parts than we'd like: 3 interactions (4, including gravity), each with its own coupling parameter; 6x3 fermion multiplets; and either one (or more) Higgs doublets, or some more complicated dynamics that produces equivalent effects at low energy.   

In this accounting, only the enumeration of fermions requires further explanation.   It is not appropriate, in a gauge theory, to regard particles related by gauge symmetry as truly independent entities.  Thus the left-handed up and down quarks (related by electroweak $SU(2)$) of all three colors should be counted as one multiplet; the right-handed up quarks of all colors and the right-handed down quarks as two more; the left-handed electron and its neutrino as another; and the right-handed electron and neutrino as two others.   The different families are not related by gauge symmetry, at least within the Standard Model, so their corresponding multiplets represent independent degrees of freedom.    

In formulating the electroweak $U(1)$ couplings of fermions, phenomenology requires some peculiar-looking choices of fractions.   If we require closure using the known degrees of freedom in the Standard Model, and demand absence of violations of gauge symmetry not only classically, but also quantum mechanically (anomaly cancellation), the choices are actually severely constrained; still, one might hope for more luminous, less arcane insight into the origin of those numbers.   

\subsection{Loose Ends}

Once we look beyond its core, we find many loose-ended strands both within the Standard Model itself, and in its account of Nature:

The Higgs sector is poorly constrained theoretically and, at least for the moment, uncharted experimentally.  (See the concluding Note.)   

The astronomical dark matter is unaccounted for.

Flavor phenomena have been smoothly accommodated, but not comparably illuminated.  There is no known theoretical principle that has, for the Yukawa couplings that encodes quark and lepton masses and mixings, anything approaching the power and coherence of the gauge principle for the core interactions.    (It is amusing to note that this contrast, in different but recognizable forms, has been with us from the earliest days classical mechanics.  Kinetic energies are simple and geometrical; potential energies are a never-ending work in progress.)  The prediction of the third generation, and the brilliant success of Cabibbo-Kobayashi-Maskawa (CKM)  phenomenology, are great achievements.  But the {\it raison d'$\hat{e}$tre\/} of family replication remains elusive, and the proliferation of theoretically unconstrained mass and mixing parameters is an embarrassment. 

Within the complicated, opaque pattern of quark and lepton masses and mixings, a few features are so pronounced as to deserve {\it qualitative\/} explanation.  These include:
\begin{description}
\item[Neutrino masses] are very small, but (in at least two cases) not zero.
\item[The $\theta$ parameter] of QCD, which might introduce $T$ violation into the strong interaction, is very small: $|\theta | \lesssim10^{-10}$.
\item[The top quark mass] $M_t \approx 172$ GeV is much larger than the mass of any other quark or lepton.  It is the only fermion for which the Yukawa coupling to the electroweak condensate, and presumably to the Higgs particle, is of order unity.  In so far as the other couplings are small, their precise value might be complicated to calculate from fundamentals, but for the top quark there is less excuse.   
\item[The electron mass], and also the masses $m_u, m_d$ of the light quarks, correspond to tiny Yukawa coupings, $y_{e, u, d}  \sim 10^{-5}-10^{-6}$.   
\end{description}
The first two of these features have been connected to promising, though still speculative, fundamental ideas, as I'll elaborate below.   

\subsection{Gravity}

Gravity, in the form of general relativity, can be brought into the standard model smoothly.  Both quantitatively and qualitatively, however, the union is inharmonious.  

Quantitatively: The observed strength of gravity -- technically, the coefficient of the Einstein-Hilbert term -- introduces a new large mass scale, the Planck mass $M_{\rm Planck} \sim 10^{18}$ GeV.    This is far larger than any other mass scale within the standard model.  That disparity defines a {\it first hierarchy problem}.  The observed cosmological term -- technically, the coefficient of a pure volume term (and thus a universal pressure)
\begin{eqnarray}
\Delta L ~&=&~  \lambda \int \sqrt g \, d^4x \nonumber \\
(\lambda )^{\frac{1}{4}} ~&\approx&~ 2 \times 10^{-3} \, {\rm eV}
\end{eqnarray}
introduces a mass scale that is significantly smaller than most standard model mass scales, though it is comparable to probable neutrino masses.   That disparity defines a {\it second hierarchy problem}. 

QCD, through asymptotic freedom, advances the first hierarchy problem significantly.   We can ask, specifically, why there is a big disparity between the QCD scale and the Planck mass.   If we extrapolate logarithmic running of the coupling all the way to the Planck scale, we find that the QCD coupling is not terribly small there.  (See immediately below, for a closely related discussion of coupling constant unification.)   Reading it the other way: If we hypothesize a ``generic, but not large'' effective coupling at the Planck scale, then we will compute the proton mass to be many orders of magnitude smaller than the Planck mass, as is observed.   Here of course I've assumed that the {\it quark\/} masses are negligibly small, as they are in Nature.  We don't know why, so the first hierarchy problem is not completely solved.   

The second hierarchy problem will be discussed extensively in the cosmology session of this meeting.   

Qualitatively: The minimal implementation of general relativity within the standard model gives us an elegant, successful working theory of quantum gravity (modulo the hierarchy problems), but that minimal theory has not been formulated nonperturbatively, and at that level it almost certainly fails to exist.  Moreover, there appear to be fundamental issues in black hole physics and in the treatment of cosmological singularities, that the minimal theory cannot address even qualitatively.    Those problems will be discussed extensively in the string session of this meeting.

\bigskip

%%%%%%%%%%%%%

\section{Approaches: ``Modest'' Improvements}

%%%%%%%%%%%%%%%%%%%%%

\bigskip

\subsection{Unification and Supersymmetry}

At the level of quantum numbers, the interactions and multiplets of the Standard Model fit beautifully into a unified theory.  The most attractive unification is based on the spinor representation of $SO(10)$, though variants are certainly possible, and have their advocates. 

The renormalization group shows us how observed (low-energy) couplings might diverge from the (high-energy) equality that nonabelian symmetry requires.  

Famously, if we construct our unified theory using only the degrees of freedom in the standard model, the resulting constraint among observed couplings doesn't quite work, 
while if we extend the theory to include the degrees of freedom required for approximate supersymmetry, at masses $\sim 10^2-10^4$ GeV, we get a much more successful relation.  The observed low-energy couplings extrapolate to a common value $g_U$ at a large mass scale $M_U$, with
\begin{eqnarray}
g_U ~&\approx&~ .7 \ \ \ (\alpha_U \approx \frac{1}{25}) \nonumber \\
M_U ~&\approx&~ 2 \times 10^{16} \, {\rm GeV}
\end{eqnarray}
There is of course some play in these numbers, which arise in the minimal implementation of low-energy supersymmetry.   Taking them at face value, however, we find many good features:
\begin{description}
\item [The unified coupling is not terribly strong.]  Thus the extrapolation of weak-coupling formulas for running of the couplings is internally consistent.
\item [The unified coupling is not terribly weak.]  Thus $M_U$ is plausibly associated with a scale for dynamical symmetry breaking.  (Which after all is what it is!)   
\item [The unified scale is not too large.]  Thus grave uncertainties associated with quantum gravity at the Planck scale are sequestered.
\item [The unified scale is not too small.]  Proton decay is sufficiently suppressed.  More accurately, the calculable contribution to proton decay due to exchange of the new gauge bosons introduced by unification is sufficiently suppressed.   Other less universal contributions, especially the contribution due to colored Higgsino exchange, are potentially dangerous, and constrain model-building.  
\item [The unified scale provides reasonable, though not perfect, input to a neutrino seesaw:] $m_\nu \sim \frac{M_t^2}{M_U}$ 
\item [One finds remarkably good, though not perfect, unification with gravity:] $M_U \sim M_{\rm Planck}$.  Since gravity is directly sensitive to energy it runs as a power when probed at high virtuality, even classically, rather than merely logarithmically due to quantum vacuum polarization, as for the gauge couplings.   Remarkably: If we expand the running of couplings calculation to include gravity, we find approximate unification, even though gravity is {\it abysmally\/} weaker -- by a factor $\sim 10^{-40}$! --  than the other forces (measured, of course, at the level of elementary particles) at all practically accessible distances and energies.  
\end{description}

To me the unforced fit of fermion families into spinor multiplets together with the accurate, multi-advantaged unification of couplings render unification, as quantitatively enabled by low energy supersymmetry, into a most compelling speculation.   

Detailed implementations of low energy supersymmetry typically include a lightish mimic of the minimal standard model Higgs particle.  A second doublet is mandatory, as well.  Low energy supersymmetry also produces, in many but not all implementations, a dark matter candidate.  

For all its attractiveness low energy supersymmetry both poses many theoretical challenges, and faces many experimental challenges.  There is no consensus on the mechanism of supersymmetry breaking, and no existing mechanism seems entirely satisfactory.   There is no evidence, so far, for any of the many additional contributions to flavor-changing processes, or $T$-violating processes, that low energy supersymmetry brings in.   There is no reliable encouragement, so far, for supersymmetric dark matter candidates.   And of course, most importantly, there is no evidence so far for any superpartners!  In anticipation mixed with dread, we await the verdict of Nature.   

Supersymmetry is, of course, a symmetry principle.  It extends Poincare symmetry to include transformations into quantum dimensions, described by anticommuting coordinates.  
Two other popular symmetry-based suggestions for going beyond the Standard Model, that address some of the same questions as supersymmetry (but so far lack, as far as I know, comparable success stories), are technicolor and extra classical dimensions.  They are based on additional gauge symmetry and on a more conventional extension of Poincare symmetry, respectively.

\subsection{$\Theta$ Problem and Axions}

The theory of the strong interaction (QCD) admits a parameter, $\theta$, that is observed to be unnaturally small: $|\theta | < 10^{-9}$.   That suspicious ``coincidence'' can be understood by promoting translation of $\theta$ to an asymptotic or classical quasi-symmetry, that is spontaneously broken.

The axion field $a$ is established at the Peccei-Quinn transition, when a complex order-parameter field $\phi$ acquires an expectation value $F$:
\begin{equation}
\langle \phi \rangle ~=~ F e^{i\theta} ~=~ F e^{ia/F}
\end{equation}

At the transition, which occurs (if at all) in the very early universe, the energy associated with varying $\theta$ is negligible, and differences from the minimum $\theta =0$ can be imprinted.  They store field energy that eventually materializes, with density roughly proportional to $F\sin^2 \theta_0$ today.  

If no inflation occurs after the Peccei-Quinn transition then the correlation length, which by causality was no larger than the horizon when the transition occurred, corresponds to a very small length in the present universe.  To describe the present universe on cosmological scales, therefore, we should average over $\sin^2 \theta_0$.  One finds that $F \sim 10^{12}$ GeV corresponds to the observed dark matter density. 

Since experimental constraints require $F \geq 10^{10}$ GeV, axions are almost forced to be an important component of the astronomical dark matter, if they exist at all.  So it seems interesting to entertain the hypothesis that axions provide the bulk of the dark matter, and $F \cong 10^{12}$ GeV.  That has traditionally been regarded as the default axion cosmology.  A cosmic axion background with $ F \cong 10^{12}$ GeV might be detectable, in difficult experiments.  Searches are ongoing, based on the conversion $a \rightarrow \gamma \gamma (B)$ of axions into microwave photons in the presence of a magnetic field.    

If inflation occurs after the Peccei-Quinn transition, things are very different.  In that scenario a tiny volume, which was highly correlated at the transition, inflates to include the entire presently observed universe.  So we shouldn't average.  $F > 10^{12}$ GeV can be accommodated, by allowing ``atypically'' small $\sin^2 \theta_0$.  

But now we must ask, by what measure should we judge what is ``atypical''?  In the large-$F$ scenario, most of the multiverse is overwhelmingly axion-dominated, and inhospitable for the emergence of complex structure, let alone observers.  Thus it is appropriate, and I would say necessary, to consider selection effects.  

Fortunately, while $\theta_0$ controls the dark matter density, it has little or no effect on anything else.  It is hard to imagine a clearer, cleaner case for applying anthropic reasoning.  The result of such an analysis is encouraging.  Taken at face value, it suggests that in the large $F$ axion cosmology the typical observer sees a ratio of dark to baryonic matter close to what we observe in our neighborhood (that is, in the universe visible to us!).

Very recently Arvanitaki and Dubovsky, elaborating earlier work by themselves and others, have argued that axions whose Compton wavelength is a small multiple of the horizon size of a spinning black hole will form an atmosphere around that hole, populated by super-radiance.  That atmosphere can affect the gravitational wave and x-ray signals emitted from such holes, possibly in spectacular ways.  Since 
\begin{eqnarray}
(m_a)^{-1} ~&\approx&~ 2 \, {\rm cm.} \ \frac{F}{10^{12} \, {\rm GeV}} \nonumber \\
R_{\rm Schwarzschild} ~&\approx&~ 2 \, {\rm km}. \, \ \ \frac{M}{\, M_{\rm Sun}}
\end{eqnarray}
this provides a promising window through which to view $F \geq 10^{15}$ GeV axions. 

\bigskip

%%%%%%%%%%%%%

\section{Experimental Frontiers}

%%%%%%%%%%%%%%%%%%%%%

\bigskip

The central glory of particle physics has been its success in describing empirical facts using beautiful ideas.  Can we keep it up, on the empirical side?  

The LHC will be a beacon for investigating the mechanism of electroweak symmetry breaking, the possibility of low-energy supersymmetry, and a large class of dark matter candidates.   Among more speculative prospects, I'd like to mention especially the possible existence of hidden sectors, i.e. standard model singlets.   That possibility has, at least, the negative virtue of not endangering the observed quantitative accuracy of the standard model.  Of course axions and the right-handed neutrinos of $SO(10)$ are standard model singlets, but the LHC is sensitive to other types, for example scalars that mix with the Higgs particle.   

Barring discovery of some qualitatively new acceleration mechanism, accelerators at the high-energy frontier will continue to be engineering projects on an industrial scale.   Whether there will be successors to the LHC is a question where politics and economics loom large.   It's our proper duty, in any case, to integrate the fruit of our work into our surrounding culture; to keep our enterprise going strong, we'll need to do it well.   

In the pioneering days of particle physics it was cosmic rays that gave access to the highest energies.   Cosmic rays have, of course, many disadvantages and limitations as compared with accelerators, but they still do offer access to the highest energies, and perhaps satellite ``whole Earth'' monitoring can compensate for abysmal rates.   In considering unconventional accelerator technologies, it may be worth keeping in mind the possibility of producing a terrestrial version of cosmic rays -- that is, to sacrifice on intensity, focus, and energy resolution for the sake of cheapness and raw energy.   

Other fundamental issues call for quite different kinds of experiments.   The dramatic progress of atomic physics in recent years should empower new sensitive searches, through elementary electric moments, for qualitatively new sources of $T$ violation.   Beautiful work has been done to look for feebly coupled light particles (``fifth forces''); this line should be pursued vigorously, and if possible extended to the monopole-dipole case, which remains nearly virgin territory.   Finally, it is difficult to overstate the importance of the search for nucleon instability.  Along with small neutrino masses, this is the second classic signature of unification.  It is overdue.  If found, it would open a unique window on the world at $10^{-29}$ cm.    %%%%%%%

\bigskip

%%%%%%%%%%%%%

\section{Cosmic Questions: Way Beyond the Standard Model}

%%%%%%%%%%%%%%%%%%%%%

\bigskip

(Since string theory has its own session, I'm avoiding that subject.  Since cosmology has its own session, I'm mostly avoiding that, too.) 

This final part will consist mostly of questions, though some of them are leading.     

\subsection{Kinematics and Dynamics}

Quantum mechanics, like classical mechanics, is more accurately portrayed as a framework rather than as a world-model.   Yet the kinematic structure of quantum theory is both considerably more elaborate and considerably tighter than that of classical mechanics.   This suggests, perhaps, that the kinematic structure is not independent of  the dynamics.  In any case, this is a sort of ``unification'' that would seem to me desirable, though it is rarely discussed.   More concretely:

In our fundamental theories of physics we postulate commutation relations several times, in ways that seem only tenuously related, conceptually.  We postulate the Heisenberg group for quantum kinematics, and separate symmetry groups for space-time transformations and for gauge structures.  Are those groups destined to remain a direct product?  

\subsection{Dynamics and Initial Values}

Can we transcend the distinction between laws of nature and initial values?  

Let me remind you that many questions that were once existential have been subsumed into dynamics, and that this represents profound progress in our understanding of the world.   Thus we now have {\it a priori\/} accounts (given the standard model) of what are the possible chemical elements, molecules, and nuclear isotopes, not to mention the hadron spectrum.   Since the emerging standard model of cosmology needs only a few numerical inputs, it may not be silly to suspect that the ``initial value'' of the Universe might be deeply simple, and from there it's a small step to conjecture that it might not be independent of the laws of nature.   

Is there a unique ``wave function of the universe''?   If so, what makes it unique?  And if so, why does reality look so messy? 

Hartle and Hawking have made a relevant proposal, the ``no-boundary'' prescription, and greatly advanced the dialogue on this issue, though their underlying microphysical framework, i.e. Euclidean quantum gravity, may be questionable.   (Hartle reviewed that work nicely later in the conference.)   

 A plausible answer to that third question, at least, seems at hand.  Reality will inevitably look messy {\it to us}, even if the total wave function of the Universe is completely symmetric and deeply principled, because we effectively sample only a small part of it, having decohered from the rest.  

Once we begin to question the primacy of the initial value problem, it is hard to avoid asking:

Is it always appropriate to think from past, or present, to future?  Might it be the future that's somehow profoundly simple instead, or as well?   Or might time support interesting topology, with branchings and loops (enabled by singularities, or quantum fuzziness)?   

Issues of a similar kind arise, in a special form, in the theory of eternal inflation.

\subsection{The Ubiquity of Spinors}

Spinors, like particle physics, are also almost exactly 100 years old (!), having been discovered by $\acute{{\rm E}}$lie Cartan in 1913.   It is astonishing to realize that such a fundamental mathematical phenomenon, inherent in the nature of things and, according to modern understanding, at the very heart of Euclidean geometry, could have escaped human notice for so long.  (One does find premonitions in Hamilton's quaternionic treatment of rotations, and possibly earler.)  Today spinors appear prominently at many frontiers of thought: in the description of space-time fermions, as the organization of standard model families and their unification, as supersymmetry generators, in the simplest nonabelian quantum statistics, and in the theory of quantum error correction.   

{\it Why\/} are spinors so ubiquitous? Could the appearance of this common ingredient within such superficially diverse structures be hinting at new possibilities for unification?

\subsection{Information as Foundation?}

The primary ingredients of today's physical theory have been distilled and refined over a long process of coming to terms with the strange world that Nature presents to us.   From real numbers to derivatives to operators in Hilbert space, they are often extremely sophisticated concepts, from an axiomatic perspective.   One might hope to build the ultimate description of Nature on more logically primitive, less artificial foundations.  To be specific:

Is information a foundational {\it physical\/} concept?  

There are, I think, significant hints that it should be.  {\it Action\/} is central to our formulation of fundamental physics, including both the standard model and most of its speculative extensions, using path integrals.  With an assist from Planck's constant action becomes a pure number, as does entropy (with an assist, if you like, from Boltzmann's constant).  Entropy, we understand, is deeply connected to (negative) information.   Indeed Shannon's information theory bears, at several points, an uncanny resemblance to statistical mechanics and thermodynamics.   In some contexts -- most impressively, in black hole physics -- the physical action can be interpreted as an entropy.   

These analogies have been noted for decades.  But they seemed to have limited promise, to me at least, because on the information side there did not appear to be richness of structure comparable to what we need on the physical side.   Recent developments in quantum information theory have, however, unveiled a wealth of beautiful structure, and investigations of geometric or ``entanglement entropy'' in quantum field theory have established profound, natural connections to just that structure.   So it may not be crazy to hope that we might go the other way, constraining or eventually deriving the physical action from hypotheses on entanglement entropy.    

\bigskip 

{\bf Note}:  My report also contained some brief remarks on issues related to the origin of mass within the standard model, and the implications of (tentative) Higgs particle phenomenology.   Since that subject is especially topical now, and in flux, I decided to expand those brief remarks into a separate paper.   

%{\it Acknowledgements}
\section*{Acknowledgements}
This work is supported by the U.S. Department of Energy under cooperative research agreement Contract Number DE-FG02-05ER41360.

%\begin{thebibliography}{99}

%\bibitem{shapereWilczek} A. Shapere and F. Wilczek, paper in preparation.

%\bibitem{a-hetal} See also A. Arkani-Hamed, H. Cheng, M. Luty, S. Mukohyama {\it JHEP} {\bf 0405} (05): 074 (2004).

%\bibitem{kuramotoModel} J. Acebr\'on, L. Bonilla, C. Perez Vincente, F. Ritort, R. Spigler, {\it Rev. Mod. Phys}. {\bf 77} 137-185 (2005). 

%\end{thebibliography}
 
\end{document}